\documentstyle[aps,epsf,multicol]{revtex} 

\renewcommand{\v}[1]{{\bf {#1}}} 
 
\renewcommand{\sp}[2]{#1\! \cdot \!  #2}

\newcommand{\s}[1]{\mbox{sign}(#1)} 
 
\renewcommand{\d}[1]{d#1} 
\newcommand{\dd}[2] {\frac{d#1}{d#2}} 
\newcommand{\pdd}[2] {\frac{\partial#1}{\partial#2}}

\newcommand{\erf}{\mbox{erf}}

\newcommand{\vw }{\v{w}}

\newcommand{\x}{\v{x}} 
\renewcommand{\a }{\alpha} 
\renewcommand{\S}{\v{S}}

\newcommand{\bea}{\begin{eqnarray}} 
\newcommand{\eea}{\end{eqnarray}} 
\newcommand{\be}{\begin{equation}} 
\newcommand{\ee}{\end{equation}}

\begin{document}  
\title{Generation of unpredictable time series by a  
Neural Network}   
\author{Richard Metzler and Wolfgang Kinzel}
\address{Institut f\"{u}r Theoretische Physik, 
Universit\"{a}t W\"{u}rzburg, Am Hubland, D-97074 W\"{u}rzburg, Germany}
\author{Liat Ein-Dor and Ido Kanter}
\address{Minerva Center and Department of Physics, 
  Bar-Ilan University, Ramat Gan, 52900 Israel}

\maketitle

\begin{abstract}
A perceptron that learns the opposite of its own
output is used to generate a time series. We analyse 
properties of the weight vector and the generated 
sequence, like the cycle length and the probability distribution of
generated sequences.
A remarkable suppression of the autocorrelation function
is explained, and connections to the Bernasconi 
model are discussed. If a continuous transfer 
function is used, the system displays chaotic and
intermittent behaviour, with the product of the learning rate
and amplification as a control parameter.  
\end{abstract}

\begin{multicols}{2}
For every prediction algorithm that maps a
binary time series onto a binary output there is a sequence
for which it gives $100\%$ wrong predictions \cite{Kinzel:Seq.}. 
This sequence can be constructed easily by having the
algorithm predict the next bit in the time series and
continuing the series with the opposite of the prediction.
Of course, this sequence will only make one given algorithm 
with one given set of initial parameters fail completely. 
However, it is still interesting to compare
the properties of such an antipredictable sequence with one
that can be predicted with good success by the same algorithm.
We will study the statistical properties
of the time series generated by one particular
prediction machine, namely the perceptron using the Hebb 
learning rule.

The perceptron is the simplest type of feed-forward
neural networks \cite{Hertz:NeuralComp}. It consists of $N$ input units
that are connected to one output unit by $N$ synaptic
weights $w_i$, $i=1,..,N$. An input vector
$\x = (x_1,..,x_N)$ is mapped onto an output $\sigma$
by a sigmoidal function of the scalar product of $\x$ and 
$\vw$: $\sigma = f(\sum_i^N x_i w_i)$, where $f(x)=\s{x}$
is used for the so-called simple perceptron, and the error
functon $f(x) = \erf(\beta x)$ 
or the hyberbolic tangent $\tanh(\beta x)$ with an
adjustable amplification $\beta$ are common choices 
for the continuous perceptron.  In Section \ref{SEC-confus}
the sequence generated by a simple perceptron that learns the opposite
of its own output is examined, whereas in Section \ref{SEC-contin}
a continuous perceptron is used, and the differences
between the two cases are highlighted.

\section{The Confused Bit Generator}
\label{SEC-confus}
Perceptrons have been used for generating binary
time series in a simple iteration that was named
Bit Generator (BG)
\cite{Eisenstein:BG,Schroeder:Diss.,Schroeder:Cycles}:
the pattern $\x^t$ at time $t$ is an $N$-bit window of
a binary time series $\S$, $\x^t = (S^{t},\ldots,S^{t-N+1})$,
$S^t \in \{-1,1\}$.
The series is generated by the output of the perceptron:
$S^{t+1}  = \s{\sp{\x^t}{\vw}}$. For fixed $\vw$,
the sequence relaxes into a limit cycle whose 
average length increases more slowly than exponentially
with $N$. Short cycles with a length $l<2N$ are more
likely than longer ones, and the Fourier spectrum
of the sequence is dominated by one frequency
which is also prominent in the weights 
\cite{Eisenstein:BG,Schroeder:Diss.}. 
The cycles can be calculated analytically if
the weights have only one Fourier component
\cite{Schroeder:Cycles}.

In \cite{Kinzel:Seq.}, a variation of the BG was 
introduced in which the next bit of the sequence
is the opposite of the perceptron's output, and
the network learns the sequence according to the
Hebb rule with a learning rate $\eta$:
\bea
S^{t+1} &=& - \s{\sp{\x^t}{\vw^t}}; \\ 
\vw^{t+1} &=& \vw^t + (\eta/N) S^{t+1} \x^t. \label{CBG-update}
\eea
We call this system Confused Bit Generator (CBG), 
because the perceptron is told that its output was
wrong no matter what it predicted.

\subsection{Dynamics of the weights} 
\label{CBG-Dynamics}
Geometrically, $\vw$ does a directed random walk on an 
$N$-dimensional cubic lattice: each component of
the learning step is $\pm \eta/N$. Thus, while the
values of the weight components $w_i$ are real numbers,
they can only take discrete values $w_i^0 \pm n\eta/N$
with $n=0,1,2,\dots$ once the initial values $w_i^0$
are chosen.

Furthermore, each learning step has a negative overlap 
with the current $\vw$, which prevents a boundless
growth of the vector.
The norm of the weight vector fluctuates around 
an equilibrium value that can be estimated by 
replacing $\x$ with a random vector whose components
have a variance of 1, taking the 
square of Eq. (\ref{CBG-update}) and applying the
usual formalism for online learning \cite{Saad:Online}:
\bea
\langle \sp{\vw^{t+1}}{\vw^{t+1}} - 
  \sp{\vw^t}{\vw^t}\ \rangle &=& 
-\frac{2 \eta}{N} \langle \sp{\x^t}{\vw^t} \s{\sp{\x^t}{\vw^t}} \rangle
 \nonumber \\
&&+ \frac{\eta^2}{N^2} \langle \sp{\x^t}{\x^t} \rangle .
\eea
Introducing a time scale $\a$ with $\d{\a} = 1/N$ and averaging
over $\x$, this becomes a deterministic differential equation
for the norm $w$ of $\vw$ in the thermodynamic limit
$N\rightarrow \infty$:
\be
\dd{w}{\a} = -\sqrt{\frac{2}{\pi}} \eta + \frac{\eta^2}{2w}.
\ee
The attractive fixed point of this equation is $w = \sqrt{\pi/8} \eta
\doteq 0.6267 \eta$.
However, using the time series generated by the perceptron
as patterns, simulations give a slightly different
value of $w \approx 0.566 \eta$, independent of $N$
(this was already observed in \cite{Kinzel:Seq.}).
Two possible violations of the assumptions for which 
the analysis in \cite{Reents:Self-averaging} 
guarantees agreement with analytical predictions must be considered:
first, the time series patterns generated by the CBG do 
now follow a uniform distribution
(see Sec. \ref{Distribution}). Second, they are not
drawn independently from the weight vector and previous
patterns. Simulations in which a perceptron was
given patterns drawn randomly from a distribution as described
in Sec. \ref{Distribution} yield a norm $w$ that is compatible
with the analytical value of $0.6267\eta$. This indicates that
temporal correlations are responsible for the deviations.

The learning 
rate $\eta$ only sets a length scale, but does not 
influence the long-term behaviour of the system.  

In a similar fashion, the autocorrelation of the
weight vector can be calculated using the assumption
of random patterns:
\be
\langle \sp{\vw^t}{\vw^{t+\tau}} \rangle = 
  w^2 \exp \left ( -\frac{4}{\pi} \frac{\tau}{N} \right ).
\label{CBG-w_autocorr}
\ee
In some cases (see Section \ref{autocorr})
it is useful to assign an individual
learning rate $\eta_i$ to each weight component $w_i$.
A short calculation shows that the mean square norm
of each weight component is proportional to its
learning rate:
\be
\langle w_i^2 \rangle = 
  \sqrt{\frac{\pi}{8}} \frac{\eta_i}{N} \sqrt{\sum_j w_j^2}.\label{w-eta}
\ee
A component with a higher learning rate thus has
a stronger influence on the output. This also 
leads to a faster decay  of the autocorrelation:
\be
\langle w_i^t w_i^{t+\tau} \rangle = 
 \frac{\sum_j \eta_j}{\eta_i} \frac{\pi}{4} 
  \exp \left ( -\frac{\eta_i}{\sum_j \eta_j} \frac{4}{\pi} \tau \right ).
\ee
The dynamics of the weights can be linked to the 
the autocorrelation 
function $C_j^t$ of the sequence, defined by
\be 
C_j^t = \sum_{i=1}^{t} S^i S^{i-j},
\ee
where $t$ is the number of patterns summed over.
Simply add $t$ update steps according to Eq. (\ref{CBG-update}):
\be
w_j^t  = w_j^0 + \sum_{i=1}^{t} (\eta/N) S^i S^{i-j}
 = w_j^0 + (\eta/N) C_j^t. \label{CBG-w_corr}
\ee
Each value $C_j^t$ for $1\leq j \leq N$ corresponds
to the distance of the weight vector from its starting
point along one axis in the $N$-dimensional weight space,
measured in units of $\eta/N$. This point is important 
and will be exploited in the following paragraphs.

\subsection{Cycles and transients}
\label{CBG-cycles}
The CBG is a deterministic map with a discrete, finite 
state space. This means
that it falls into a cycle of some length $l$
eventually:  both sequence and weights repeat
after $l$ steps, i.e. $\vw^t = \vw^{t+l}$ or alternatively
$C_j^l = 0$ for  $1\leq j \leq N$ after $l$ steps.
This means that $l$ must be divisible by 4, since only
sequences with $l \bmod 4 =0$ can have an autocorrrelation
of 0. Also, a lower bound for $l$ can be given: 
for the $l$th autocorrelation value one gets $C_l^l = l \neq 0$,
therefore $l>N$. By renaming indices, one finds
for a periodic sequence $C_j^l = C_{l-j}^l$.
If $l\leq 2 N$, one thus gets $C_j^l \equiv 0 $ for all
$j < l$. In Ref. \cite{Mertens:Bernasconi}, it is conjectured that 
such a sequence does not exist except for any $l$ except $l=4$.
If this is true, $l>2 N$ must hold for $N >3$.

An upper bound on the cycle length can be found by 
estimating how many states in weight space the weight 
vector can take. Assuming that it stays inside an 
$N$-dimensional hypersphere of radius $w_f =0.566\eta$
and volume $V = w_f^N \pi^{N/2}/\Gamma(N/2 +1)$ 
, we can divide that volume by the volume of a unit cell,
$(\eta/N)^N$, and expand using Stirling's equation.
We find that the number of possible states in weight space
scales approximately like $5.45^N/\sqrt{N}$. Combining
this with $2^N$ possible sequences gives $10.9^N/\sqrt{N}$
possible states of the system.

\begin{figure}
\begin{center}
\epsfxsize= 0.95\columnwidth
  \epsffile{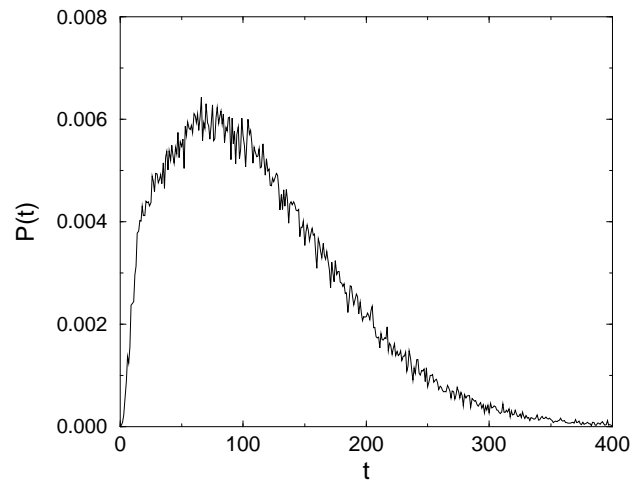}
\end{center}
\caption{Distribution of transient lengths $t$ of a CBG with 
$N=8$. The small probability of short $t$ indicates that only
a small fraction of state space is part of a cycle.} 
\label{VBG-med}
\end{figure}

Simulations show that not all of these states are part of a 
cycle: starting from random initial conditions, there is 
a transient whose median length scales approximately like 
$2.04^N$. The transient distribution (Fig. \ref{VBG-med}) 
shows that not all states have the same probability 
of being part of a cycle: the probability for a very 
short transient is smaller than that for a longer one,
which implies that some sort of annealing occurs
during the first steps.  
Simulations were done with random 
initial sequences and random initial vectors normalized 
to $w = 0.566\eta$.

\begin{figure}
\begin{minipage}[b]{0.48 \columnwidth}
    \epsfxsize= 0.99\columnwidth
    \epsffile{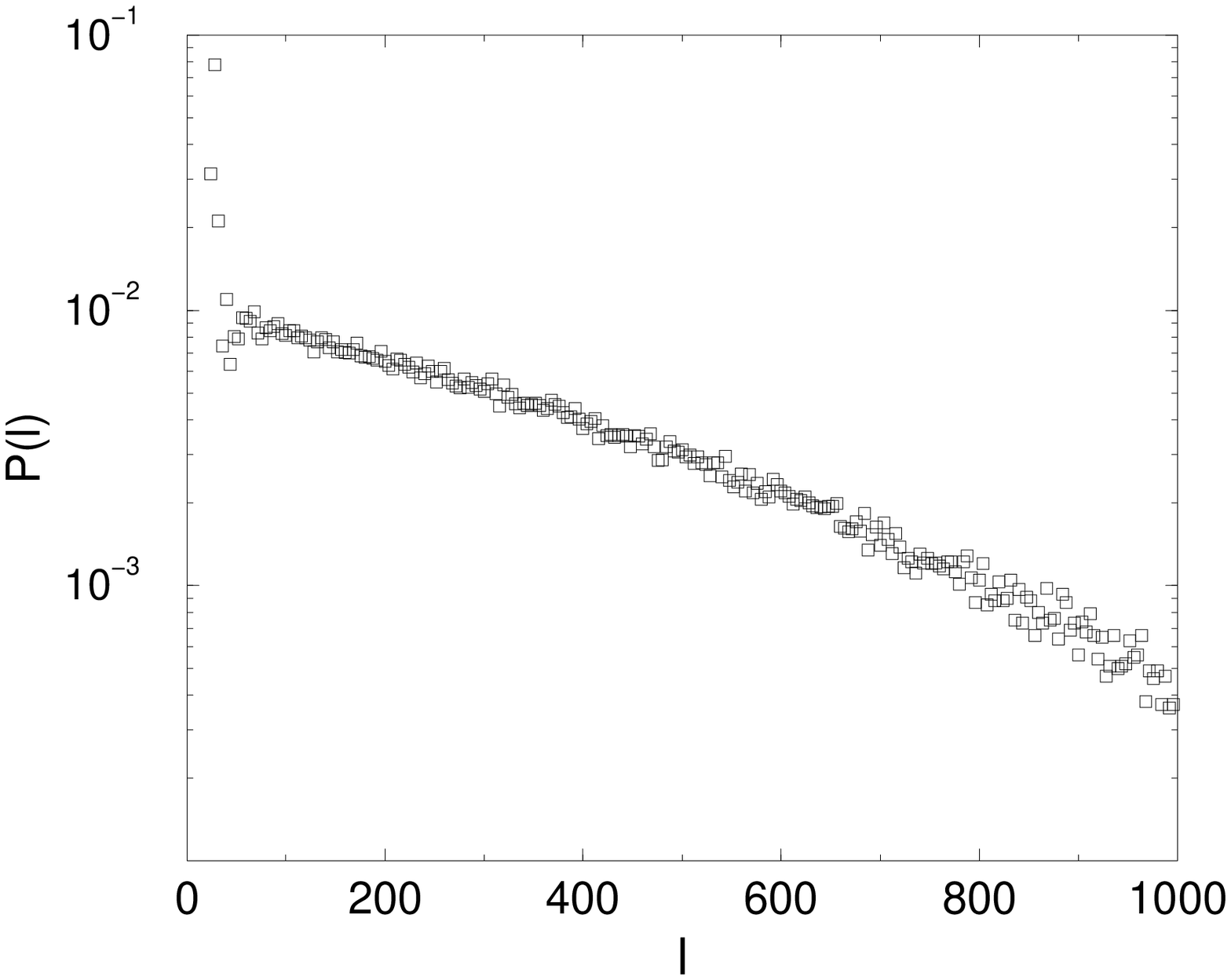}
\end{minipage}
 \begin{minipage}[b]{0.48 \columnwidth}
   \epsfxsize= 0.98\columnwidth
   \epsffile{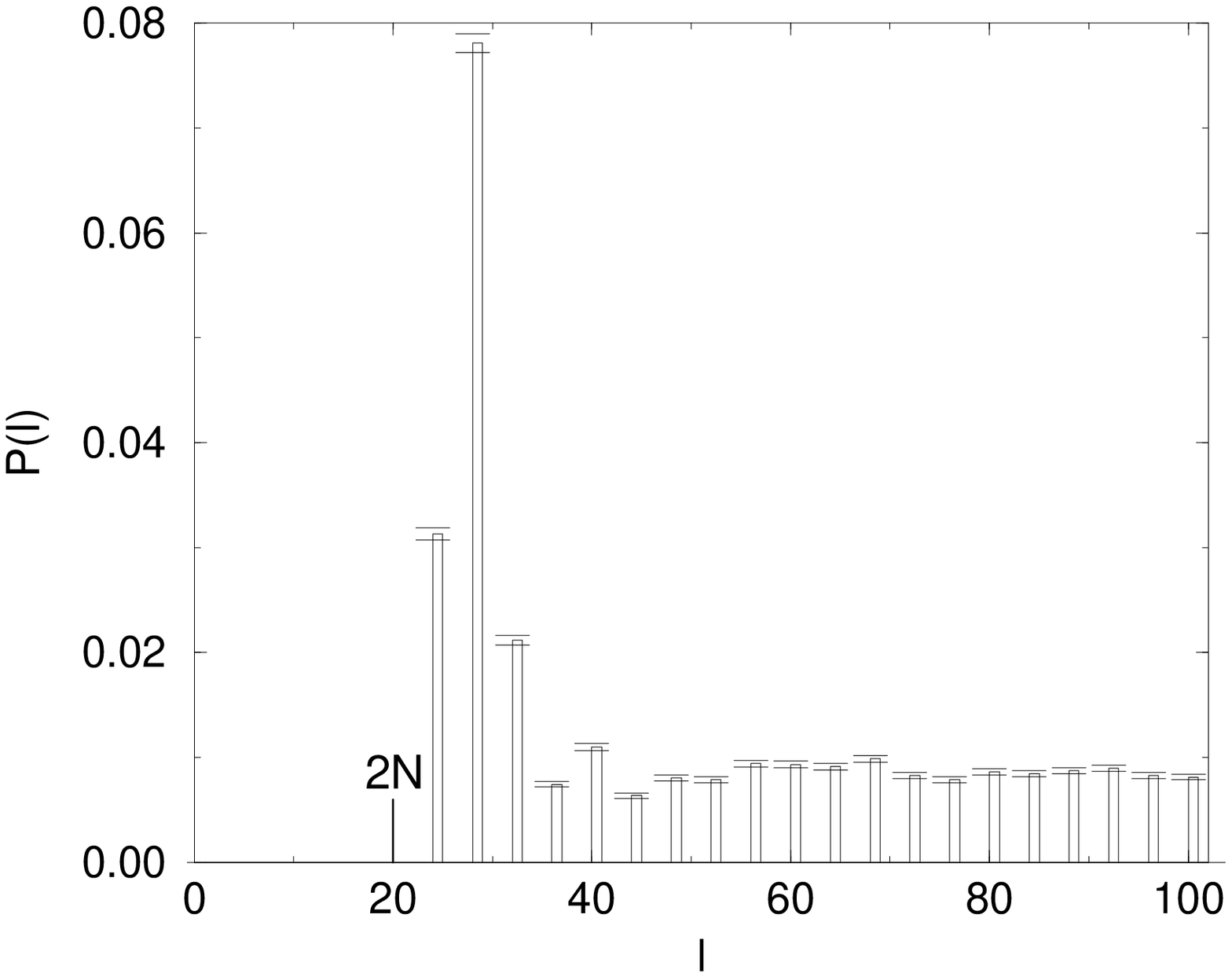}
\end{minipage}
\caption{Distribution of period lengths of a CBG with $N=10$ on 
logarithmic and linear scales. 
The initial weights and sequence were random. All $l$
are divisible by 4, and $l> 2N$. Error bars denote the
standard error.} 
\label{VBG-cycdist}
\end{figure}

\begin{figure}
\epsfxsize= 0.95\columnwidth
  \epsffile{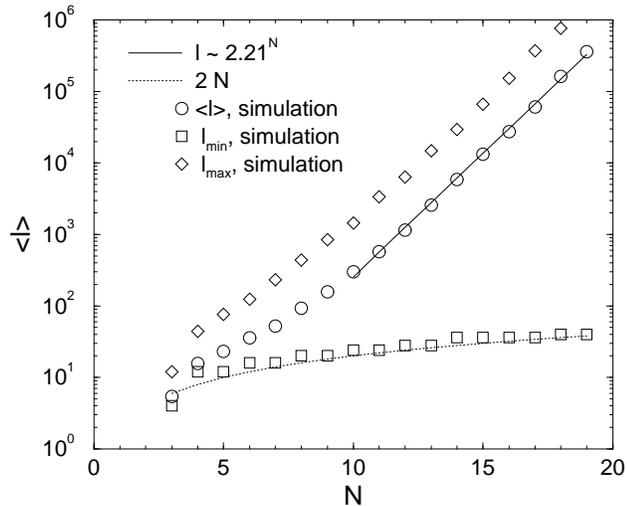}
\caption{Average, smallest and largest cycle $l$ found
in simulations with 1000 random initial conditions for each 
value of $N$. The full line is an exponential fit to the 
data for $N>9$, the dotted line denotes the theoretical 
lower bound of $l>2N$ for $N>3$.}
\label{CBG-avgl}
\end{figure}

The distribution of cycle lengths $l$ found in simulations
shows the expected features (see Fig. \ref{VBG-cycdist}): a 
minimum cycle length $l>2N$ and no cycle lengths $l$ that
are not divisible by 4. 
There is a distinct maximum  near the minimum cycle 
length and a broad distribution 
that falls off slightly faster than exponentially for large $l$.
The average of $l$ scales approximately like $2.2^N$, as
seen in Fig. \ref{CBG-avgl}.
The fact that the largest $l$ that is found scales
exponentially with $N$ suggests
that there is an exponential number of different cycles.
\end{multicols}
\begin{multicols}{2}

\subsection{Autocorrelation function and the Bernasconi model}
\label{autocorr}
\begin{figure}
\epsfxsize= 0.95\columnwidth
  \epsffile{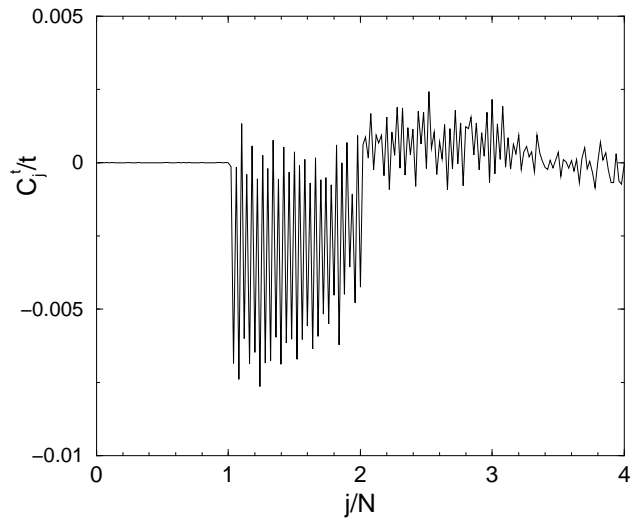}
\caption{Autocorrelation function $C_j^t/t$
of a CBG with $N=50$,
averaged over $t = 2\times 10^6$ patterns.} 
\label{VBG-corr}
\end{figure}

The autocorrelation of the sequence shows some peculiarities,
as seen in Fig. \ref{VBG-corr}:
As explained, the first $j$ values correspond to components
of $\vw$. Since $w$ is finite, $C_j^t$ is bounded for 
$1\leq j \leq N$, i.e. it does not grow like $\sqrt{t}$
as it would for a random sequence. The values for 
$N<j\leq 2N$ show negative correlations that grow 
linearly with $t$ for even $j$, whereas they are compatible
with a random sequence for odd $j$.
Between $2N$ and $3N$, correlations are positive for even 
$j$ and 0 for odd $j$.  
These effects appear for all $N$ in both the transient and the cycle
as long as the cycle length is much larger than $N$.

Bit series with low autocorrelations are of interest in 
mathematics and have applications in signal processing
\cite{Schroeder:Zahlentheorie}.
It is therefore interesting whether the CBG generates sequences
with autocorrelations significantly lower than for random series.
Two measures are commonly used in the literature:
for periodic sequences of length $l$, an energy function
(which is studied in the so-called Bernasconi model 
for periodic boundary conditions \cite{Bernasconi:LowAuto}) 
can be defined by
\be 
H_p = \sum_{j=1}^{l-1} (C_j^{l})^2 = 
  \sum_{j=1}^{l-1}(\sum_{i=1}^{l}S^i S^{i+j})^2.
\ee
Results on the ground states of this Hamiltonian can be found in 
\cite{Mertens:Bernasconi}. By trial and error, initial conditions for
the CBG can be found which yield cycles slightly larger
than $2N$, for which all value of $C_j^l$ except one are 0.
However, even for the best sequences we found, 
$ H_p$ was larger than the known ground state energies 
by at least a factor of 2.

The original model does not use 
periodic boundary conditions: in a 
sequence of length $p$, only the sum 
over $p-j$ different terms with a lag of 
$j$ can be calculated. The energy is therefore
given by
\be
H_{ap} = \sum_{j=1}^{p-1} (C_j^{p-j})^2
\ee 
(note the summation limits).
The so-called merit factor $F$ introduced by Golay \cite{Golay}
is defined by
\be
F = \frac{p^2}{2 H_{ap}}.
\label{CBG-def_F}
\ee
A merit factor of 1 is expected for a random sequence;
lower autocorrelations yield higher $F$. The theoretical
limit for large $p$ is conjectured to be about $F=12$ 
\cite{Bernasconi:LowAuto}, whereas optimization routines typically find
sequences with $5< F<9$ (see \cite{deGroot:LowAuto} and 
references therein) and exact enumeration for small $p$ suggests 
$\lim_{p \rightarrow \infty} F= 9.3$ for the optimal sequence
\cite{Mertens:Exhaust}.

\begin{figure}
\epsfxsize= 0.95\columnwidth
  \epsffile{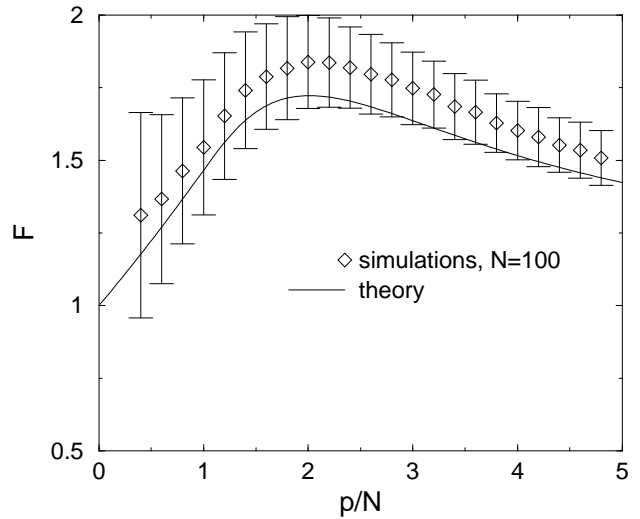}
\caption{Merit factor $F$ as a function of scaled 
sequence length $p/N$, compared to Eq. (\ref{CBG-mf2}). 
Error bars denote the
standard deviation of $F$ for $N=100$, 
not the standard error.} 
\label{CBG-meritf}
\end{figure}

To estimate the merit factor of sequences
generated by the CBG analytically, we solve Eq. (\ref{CBG-w_corr})
for ${C_j^t}^2$ and use the autocorrelation of the weights 
given by Eq. (\ref{CBG-w_autocorr}):
\bea
 \langle (C_j^{p-j})^2 \rangle &=& 
     \frac{N^2}{\eta^2} 
       \langle {w_j^0}^2 + {w_j^t}^2 - 2 w_j^t w_j^0 \rangle \nonumber \\
      &=&    \frac{\pi}{4} N 
       \left (1 - \exp \left (- \frac{4}{\pi}\frac{p-j}{N} \right) \right).   
\label{CBG-csq}
\eea
The energy can be expressed as a sum or approxated by 
an integral in continuous variables $\alpha = p/N$ and 
$\beta = j/N$. Since Eq. (\ref{CBG-csq}) only holds for
$1\leq j\leq N$, ${C_j^{p-j}}^2=p-j$ must used for $j>N$.
We get the expression
\bea
H_{ap} &=& \sum_{j=1}^{p-1} N \frac{\pi}{4} 
  \left (1- \exp \left 
      (-\frac{4}{\pi}\frac{p-j}{N} \right) \right) \nonumber \\
&\approx& \int_0^{\alpha} 
    N^2 \frac{\pi}{4} (1- \exp(-(4/\pi)(\alpha-\beta))) d\beta 
    \nonumber \\
&=& N^2 \frac{\pi}{4} (\alpha - \frac{\pi}{4}
      (1- \exp(-\frac{4}{\pi} \alpha))) \mbox{ for }j\leq N\mbox{ and }
      \label{CBG-mf1} \\
H_{ap} &=& N^2\left(\frac{\pi}{4} 
  \left (  
    1 - \frac{\pi}{4}(\exp(\frac{4}{\pi}(1-\alpha)) 
    - \exp(-\frac{4}{\pi}\alpha) ) \right ) \right.\nonumber \\
  & &+ 
 \left. \frac{1}{2}(\alpha-1)^2 \right ) \mbox{ for } j>N.
\label{CBG-mf2} 
\eea
The corresponding merit factor is compared to
simulations in Fig. \ref{CBG-meritf}: Eqs. (\ref{CBG-mf1})
and (\ref{CBG-mf2}) give qualitatively  correct results, but differ
from the observed values by roughly $10\%$. The
feedback mechanisms of the CBG cause a faster decay of $C_j^t$
than predicted for random patterns. 
We observed in Section \ref{CBG-Dynamics} that individual 
learning rates can make $C_j^t$ decay faster for some $j$
and slower for others. The search for a minimal $H_{ap}$
can be written as an optimization problem in the 
continuous function $\eta(\beta)$, where $\eta_j = \eta(j/N)$.
Solving this problem with a variational approach, one
finds that it is sensible to give the last 41\% of the
weights a learning rate and norm of zero, and increase
the learning rate continuously towards  components
with smaller indices. Unfortunately, even this optimization
does not improve the merit factor beyond $F=1.74$ in theory 
and $\langle F \rangle =1.86$ in simulations. This is
still a lot worse than the results of other optimization
methods \cite{Bernasconi:LowAuto,deGroot:LowAuto}, 
so the CBG is not a competitive generator
of low-autocorrelation sequences. Nevertheless, it 
has some interesting possibilities:

\subsection{Shaping the autocorrelation function}

\begin{figure}[t]
\epsfxsize= 0.98\columnwidth
  \epsffile{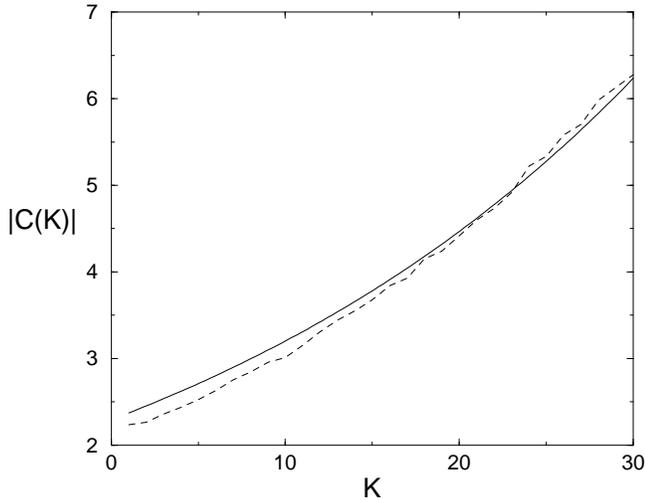}
\caption{Profile of the average absolute value of the autocorrelation function
 $|C_j|$ which was achieved by a CBG with N=30 input units 
using $\eta_j=\exp(-2j/N)$ (dashed curve). 
The solid curve stands for the desired profile, 
$|C_j|=A\,\exp(j/N)$,
 where $A=\sqrt{\pi/8}$ $ \sqrt{(e^{-2}-1)/(e^{-2/N}-1)}$.} 
\label{exponent}
\end{figure}

\begin{figure}[t]
\epsfxsize= 0.92\columnwidth
  \epsffile{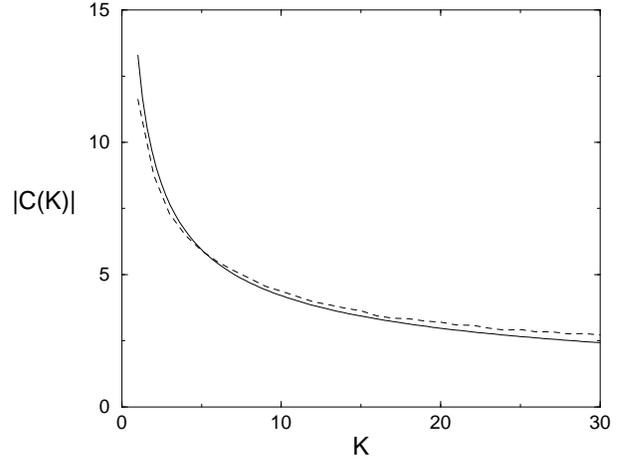}
\caption{Profile of the average absolute value of the autocorrelation function
 $|C_j|$ which was achieved by a CBG with N=30 input units 
using $\eta_j= j/N$ (dashed curve). 
The solid curve stands for the desired profile, 
$|C_j|=2.427\sqrt{j/N}$. }
\label{polinom}
\end{figure}

Being able to suppress autocorrelations, the CBG 
is also capable of controlling the 
shape of the autocorrelation function in the long-time limit.
Using Eqs. (\ref{w-eta}) and (\ref{CBG-w_corr}) in the limit where 
$\v{w}^0 \to \v{0}$ and for non-negative learning rates, 
one can obtain the inverse relation 
between the square of the autocorrelation function $(C_j^t)^2$ and the 
corresponding learning rate ${\eta_j}$ 
\be
 \langle |C_j^{t}| \rangle = 
     \sqrt{\frac{\pi}{8}} \sqrt{{\sum_{i=1}^{N}{\eta_i}}\over{\eta_j}}. 
\label{C-eta}
\ee
Thus, any desired shape of the autocorrelation function is
 achievable by using the appropriate profile for $\eta_j$ which can  
be extracted from Eq.(\ref{C-eta}).
High performance of the CBG as a producer of sequences with
 specific desired shapes of the autocorrelation function is observed
 in simulations.
This feature of the CBG is demonstrated in Figs. \ref{exponent}
and \ref{polinom}  where  both an
exponential and a polynomial profile of the autocorrelations are 
successfully generated. The slight deviations from the target 
profile are probably due 
to a violation of the assumption of random patterns.  
Simulations are done for a CBG with 30 input units. 
The autocorrelations are calculated for time windows of 100000 bits and
 are averaged over 1000 such successive windows.
Checking a wide variety of shapes, the CBG exhibits a 
decent capability of achieving  
the expected profiles. It could be used as an 
alternative mechanism for generating colored 
binary sequences using local rules instead of
nonlocal mechanisms auch as Fourier transforms.

The limited use of the CBG in generating sequences with a
high merit factor may be related to phase space arguments:
as seen in Sec. \ref{CBG-cycles}, the CBG can still generate
exponentially many different time series depending on 
initial conditions, whereas there are very few sequences 
with the highest achievable high merit factors (see \cite{Mertens:Density}
for the density of states with cyclic boundary conditions).
The mechanism of the CBG allows for manipulation of the
autocorrelation function only if the constraints on the
desired sequence are not too strong, such as suppressing
all of the elements of $C_j$ on a short time scale.
On the other hand, choosing a given shape for long-time
averages of $C_j$ still allows for many realizations of
the sequence.

\subsection{Distribution of generated sequences}
\label{Distribution}
\begin{figure}
\epsfxsize= 0.87\columnwidth
  \epsffile{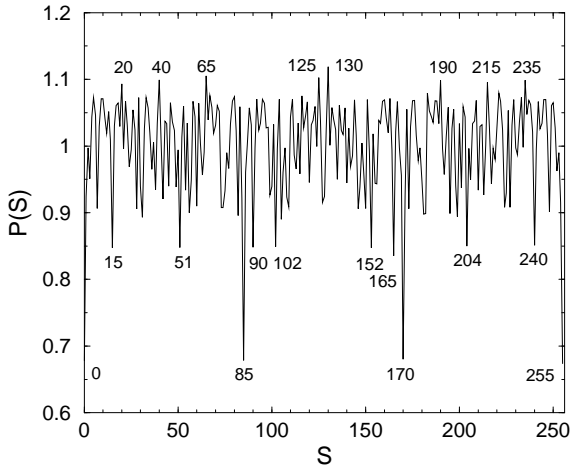}
\caption{Histogram of 8-bit subsequences generated by a
CBG with $N=50$, averaged over $5\times 10^6$ steps. 
The $y$ axis gives the probability of a subsequence normalized to 1.} 
\label{VBG-histo}
\end{figure}

The structure in $C_j$ shows that the CBG does not
generate a random sequence. This becomes more obvious
in a histogram of subsequences generated by the system.
Fig. \ref{VBG-histo} shows the probability distribution
of 8-bit substrings from a run of a CBG with $N=50$,
encoded as decimal integers. Some strings are strongly
suppressed, most notably 0 (binary 00000000), 85 (01010101),
170 (10101010) and 255 (11111111). Other sequences
with below-average likelihood also correspond to ``simple''
sequences, like 15 (00001111) and 51 (00110011). 
Continued simple sequences give high values of some
components of the autocorrelation function, which is
unlikely as explained above.

The shape of the histogram is the same for all $N$ 
in both the transient and the cycle. However,
$l$ must be much larger than the number of bins
in the histogram. The amplitude of the deviations    
from uniform distribution again goes like $1/N$.

\begin{figure}
\epsfxsize= 0.87\columnwidth
  \epsffile{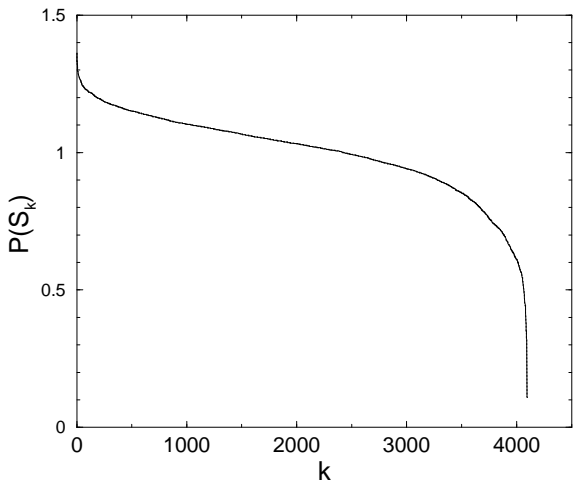}
\caption{Frequencies of 12-bit subsequences generated by a
CBG with $N=50$, ordered by rank $k$ and normalized to an 
average of $1$ (solid line). The frequencies reproduced by a Markov
process are also displayed (dotted line), but indistinguishable from
the first curve.} 
\label{VBG-historank}
\end{figure}

Ordering histograms in descending rank order often
reveals insights into the underlying processes and
phase space structure (see e.g. 
\cite{vdBroeck:Learning,Kanter:Markov}). In our 
case, the rank ordered histogram does not show a
power law or other universal behaviour, as seen 
in Fig. \ref{VBG-historank}.
 
One way to explain this histogram is by finding
the stationary distribution for a biased 
random walk on a DeBruijn graph, as was done in
\cite{Kanter:Markov,Challet:Relevance}:
a subsequence $S$ is followed by 1 with probability
$p_S$ and by 0 with probability $1-p_S$. It is 
possible to reproduce the histogram of the CBG 
accurately this way; however, one has to take the 
transition probabilities $p_S$ for each subsequence 
from simulations of the CBG. There is no obvious 
way of calculating them analytically, and taking
random transition probabilities does not 
reproduce the shape of Fig. \ref{VBG-historank}.  
The CBG may be considered the simplest case of a
sequence-generating perceptron that deterministically 
changes its direction. The sequence generated by it,
while complex, has many properties that can be understood
at least qualitatively, especially those that can be
linked to the autocorrelation function.
It is not by any standard a satisfying random bit sequence, 
and while results derived from the assumption of random 
patterns are usually qualitatively correct, the exact values have
to be modified. 

\section{The Confused Sequence Generator}
\label{SEC-contin}
The simplest generalization of the CBG to a 
continuous perceptron replaces
the sign function in Eq. (\ref{CBG-update}) by a
continuous sigmoidal function:
\bea
S^{t+1} &=& -\erf (\beta \sum_{j=1}^{N} w_j S_{t-j+1}) = 
 -\erf(\beta \sp{\vw^t}{\x^t}) \nonumber \\
&=& -\erf (\beta h^t);\label{CSG-upS}\\
\vw^{t+1} &=& \vw^{t} + \frac{\eta}{N} S^{t+1}\; \x^t,
\label{CSG-upw}
\eea
where the only new quantity is the amplification $\beta$,
and $h^t$ is an abbreviation of $\sp{\vw^t}{\x^t}$.
The generation of a time series by a continuous perceptron
with fixed weights was studied in a number of publications
\cite{Kanter:Analytical,Priel:Noisy,Priel:Long-term}, 
in which the system was named Sequence Generator
(SGen). We will call the mapping defined
by (\ref{CSG-upS}) and (\ref{CSG-upw}) 
Confused Sequence Generator (CSG). 

The SGen with fixed weights has a critical 
amplification $\beta_c$ that depends on 
$\vw$, below which $S=0$ is the attractive
fixed point. Above $\beta_c$, the zero solution
becomes repulsive, and the SGen generates a periodic
or quasiperiodic time series with an attractor  
dimension of 0 or 1 for most choices of $\vw$ and
$\beta$ \cite{Kanter:Analytical}. This attractor
is robust to noise \cite{Priel:Noisy}. The
time series displays chaotic behaviour only 
for very special choices of $\vw$ and $\beta$
(``fragile chaos'') if the transfer function is
monotonic, and for generic initial 
conditions (``robust chaos'') only if it is non-monotonic
\cite{Priel:Long-term}.
We will compare these properties to those of the
CSG.

\subsection{Mean-field solution for $w$}
\label{CSG-meanfield}
Similar to the CBG, the weight vector of the CSG 
does a directed random walk near the surface of 
a hypersphere of radius $w$.
Unlike the CBG, the length of the learning
steps depends on the magnitude of the output,
which in turn depends on $w$ and the outputs
in previous timesteps. To find an approximate 
solution to this self-consistency problem, we will first
ignore correlations between patterns and weights and treat
the patterns as random and independent.

In this approach, the inner field $h$ is a Gaussian random variable
of mean 0 and variance $w^2 S^2$, where 
$S^2 = \langle {S^t}^2 \rangle_t$ is the mean square
output of the system. 

The norm $w$ is found by taking the square of (\ref{CSG-upw}):
\be
{w^{t+1}}^2 = {w^{t}}^2 - 
 \frac{2 \eta}{N} \sp{\vw^t}{\x^t}
 \erf(\beta \sp{\x^t}{\vw^t}) + \frac{\eta^2}{N^2} 
{S^t}^2 \sp{\x^t}{\x^t},
\label {updating}
\ee
and averaging over the input patterns.
The self-overlap $\sp{\x}{\x}$ is on the average $N S^2$,
so the fixed point of $w$ is given by
\be
2 \langle h\; \erf (\beta h) \rangle = \eta S^4.
\ee  
The average on the left hand side can be evaluated
and leads to
\bea
\frac{4}{\sqrt{\pi}}\frac{\beta w^2 S^2}{\sqrt{1+ 2\beta^2 w^2 S^2}} 
&=& \eta S^4, 
\mbox{ or} \\
 \frac{\pi \beta \eta^2 S^6 + 
     \eta S^2 \sqrt{\pi}\sqrt{16 + \pi \beta^2 \eta^2 S^8}}{16 \beta} &=&w^2. 
\label{CSG-w_S}   
\eea

Let us now turn to $S^2$. The probability distribution
of $S$ itself is rather awkward, since it involves 
inverse error functions, and its slope diverges at
$S = \pm 1$. However, $S^2$ can be easily calculated
by using the distribution of $h$:
\bea
\langle S^2 \rangle &=& \int_{-\infty}^{\infty}
  \erf^2(\beta h) (2 \pi w^2 S^2)^{-1/2} \exp (-\frac{h^2}{2 w^2 S^2})\; \d{h}
  \nonumber \\
 &=& \frac{2}{\pi} \arcsin \left 
 ( \frac{2 \beta^2 w^2 S^2}{1 + 2 \beta^2 w^2 S^2} \right ). \label{CSG-S_S} 
\eea  
Plugging $w^2(\eta, \beta,S^2)$ from Eqs. (\ref{CSG-w_S}) 
into (\ref{CSG-S_S}) and solving
numerically, one obtains a self-consistent solution for $S^2$.
A closer look at the equations reveals that if a new quantity
$\gamma = \eta \beta$ is introduced, 
only $\gamma$ enters into the equation for $S^2$,
and $w^2$ is of the form $w^2 = \eta^2 \hat{w}^2(\gamma)$, so only one 
curve must be considered. This is intuitive,
since a higher $\eta$ eventually leads to a higher $w$, which
has the same effect on $S^2$ as having a smaller $w$, but 
multipying $\sp{\vw}{\x}$ with a higher factor $\beta$.  

The map defined by (\ref{CSG-upS}) and (\ref{CSG-upw}) always
has the trivial solution $S=0$. Only for a sufficiently high
$\gamma>\gamma_c$ are the outputs high enough to sustain a nonvanishing
solution.
Note that $S=0$ is always an attractive solution for
all $\gamma <\infty$, but its basin of attraction
becomes smaller for larger $\gamma$.
 
\begin{figure}
\epsfxsize= 0.98\columnwidth
  \epsffile{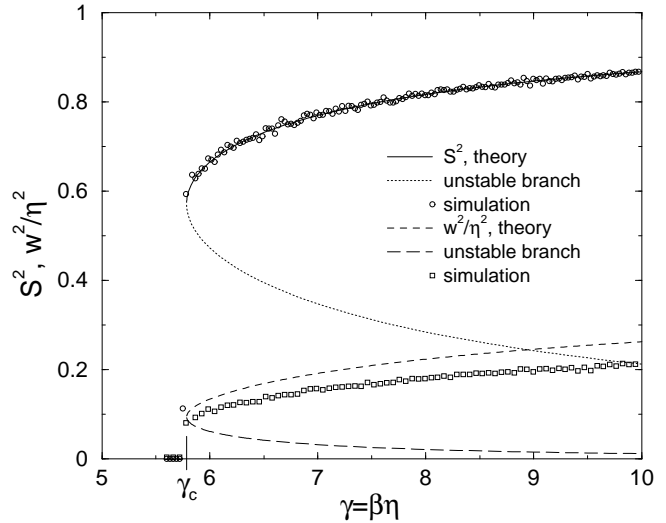}
\caption{Mean-field solution of the CBG, compared to simulations
with $N=400$.}                     
\label{CSG-S_beta}
\end{figure}

The numerical solution of Eqs. (\ref{CSG-w_S}) and (\ref{CSG-S_S})
shows that the system undergoes a saddle-node
bifurcation at $\gamma_c \doteq 5.785$, which is 
in good agreement with simulations. 
Above $\gamma_c$, two new fixed points exist, only
one of which is stable. While for $S^2(\gamma)$
excellent agreement is found between theory and simulation
(see Fig. \ref{CSG-S_beta}),
$w^2(\gamma)$ shows quantitative differences which are
caused by correlations between $\x$ and $\vw$:
the mean square overlap $\langle (\sp{\x}{\vw})^2 \rangle$
turns out to be $1.22\pm 0.01 w^2 S^2$ instead of $w^2 S^2$ 
as expected for random patterns. This causes a factor
of roughly 0.82 between the theoretical and observed 
value of $w^2$ seen in Fig. \ref{CSG-S_beta}). 
The same factor is found in the CBG. 

For large $\gamma$, $S^2$ goes to 1 (as it should,
since the system  is identical to the CBG if $\gamma = \infty$), and
the theoretical prediction for 
$w$ goes to $\sqrt{\pi/8}\eta $, just like in the CBG.

\subsection{The CSG of $m$th degree - CSG$m$}

Multi-spin interactions were studied in fields 
like neural networks \cite {multiKanter}, 
low-autocorrelated sequences \cite{Bernasconi:LowAuto} 
and error-correcting codes \cite{KanterSaad}. 
The idea to include multi-spin interactions in our 
work originated as an attempt to improve the suppression of 
the autocorrelation function achieved by the CBG. 
The existence of four-spin interactions in the Bernasconi model
implies that a CBG with multi-spin interaction might be useful
in the construction of low-autocorrelated sequences. 
However, it turns out that a CBG with
multi-spin interactions suppresses the corresponding 
multi-spin correlations instead.

In this section we apply multi-spin interactions to the CSG and
define the CSG$m$, namely a CSG in which each weight component
$w_j$ connects the corresponding input unit $x_j$ 
not only to the output but also to other $m-2$ additional input units. 

\begin{figure}
\epsfxsize= 0.98\columnwidth
  \epsffile{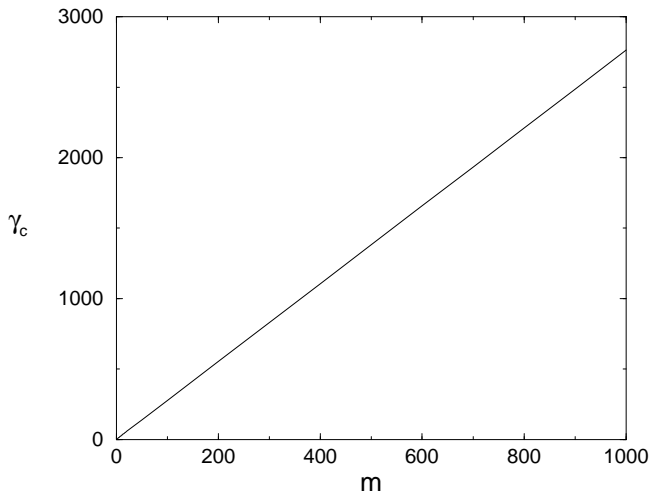}
\caption{The linear relation between $\gamma_c$ and $m$ derived from 
numerical solution of Eqs. (\ref{multi-w_S} ) and (\ref{multi-S_S}).}
\label{CSGm-m}
\end{figure}

Assigning 
$A_{j,i}$  $i=1,..,m-2$
to be the labels of the $m-2$ additional input units
 participating in the $j$th
interaction, the dynamics of a CSGm is given by:
\bea
S^{t+1} & =& -\erf (\beta \sum_{j=1}^{N} w_j S^{t+1-j}
\prod_{i=1}^{m-2} S^{t+1-A_{j,i}});
\label {multi-upS} \\
w_j^{t+1} &=& w_j^{t} + \frac{\eta}{N} S^{t+1} S^{t+1-j}
\prod_{i=1}^{m-2} S^{t+1-A_{j,i}}.
\label {multi-upw}
\eea
A similar calculation under the same assumptions which are 
used to yield the solution of the original 
CSG gives the following general set of equations:
\bea
w^2 &= &\frac{\pi \beta \eta^2 S^{2(m+1)} + 
    \eta S^2 \sqrt{\pi}\sqrt{16 + \pi \beta^2 \eta^2 S^{4m}}}{16 \beta};
\label {multi-w_S}
\\   
\langle S^2 \rangle &=& \frac{2}{\pi} \arcsin \left 
 ( \frac{2 \beta^2 w^2 S^{2(m-1)}}{1 + 2 \beta^2 w^2 S^{2(m-1)}} \right ). 
\label {multi-S_S} 
\eea
Solving numerically (\ref{multi-w_S}) and (\ref {multi-S_S}) for 
a large range of $m$ values, 
both the bifurcation point, $\gamma_c$, and the first
 non-zero values of $S^2$ and $w^2$ were found to increase 
with $m$. For $m\to\infty$ one can easily show that $\gamma_c\to \infty$
while for the non-vanishing solution $w^2\to 1$ and $S^2\to 1$ . 
Aiming to study the asymptotic behavior of $S^2$ and $\gamma_c$
in the large $m$ limit,
we set $S^2=1-\epsilon$ and find out that $\epsilon$ must decay to zero 
at least as $1/m$ in order for a non-zero solution to exist. This inverse 
relation between $\epsilon$ and $m$ derives $S^m$ terms, since 
$(1-\epsilon)^m\to 0$ unless $\epsilon<~\frac{1}{m}$. 
Inserting $S^2=1-\epsilon$ in Eqs. (\ref{multi-w_S}) and 
(\ref{multi-S_S}) 
and expanding the resulting expression  
to a power series in $\epsilon$, the inverse relation between 
$\epsilon$ and $m$ leads to linear increment of
$\gamma_c$ as a function of $m$. The numeric solutions of the system
in the large m regime supports the linear behavior of $\gamma_c$ as
derived from the aforementioned analysis (Fig. \ref{CSGm-m}).

\begin{figure}
\epsfxsize= 0.98\columnwidth
  \epsffile{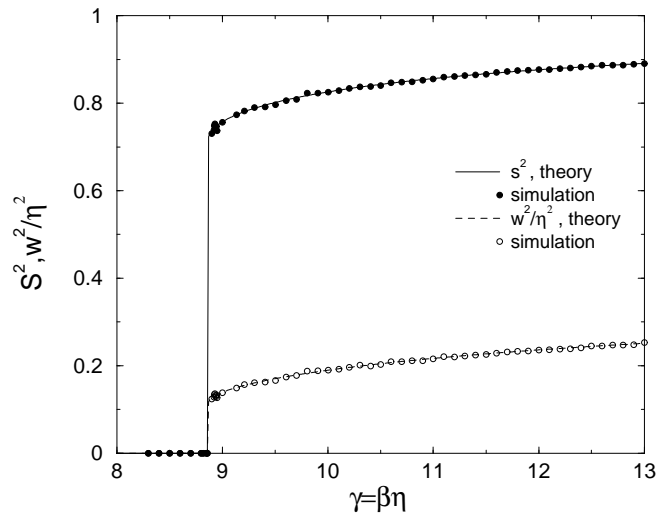}
\caption{Mean-field solution of the CSG3, compared to simulations
with $N=2000$.}                     
\label{CSG3}
\end{figure}

Fig. \ref{CSG3}  describes the numerical solution
 with respect to the simulation 
results for a system with $m=3$. This harmony between analytics and 
simulations is observed for larger $m$ as well.    

\subsection{Autocorrelation function}
The relation Eq. (\ref{CBG-w_corr}) that links
the autocorrelation function to the weights
still holds for the CSG. Since the weight
vector is bounded in the CSG as well, the same
argument can be given for the first suppression
of the first $N$ values of the autocorrelation
function. Correspondingly, $C_j^p/p$ is almost 
indistinguishable from that of the CBG
shown in Fig. \ref{VBG-corr}.

\subsection{Cycles and attractors}
\label{CBG-cyc}
The CSG can be seen as a nonlinear mapping that 
maps the vector $\x^t \oplus \vw^t$ onto $\x^{t+1} \oplus
\vw^{t+1}$. This is in contrast to previous work on
the SGen \cite{Priel:Long-term}, where the weights
were fixed and could be considered parameters of the
model rather than dynamic variables. The only real 
control parameter of this mapping is $\gamma$.

Since both the sequence and the weights now live in
a high-dimensional space of real numbers, the CBG
can display a wide variety of behaviours, depending
on $N$ and $\gamma$:

For $\gamma < \gamma_c$, the zero solution is the
only attractor, and the system will quickly 
reach $\x^t =\v{0}$ and stop developing.

For $\gamma$ slightly above $\gamma_c$, an
irregular-looking time series with the statistical
properties calculated in Section \ref{CSG-meanfield}
and displayed in Fig. (\ref{CSG-S_beta}) is generated.
However, the zero solution is still attractive, and
after some time the system will drift close to it
and stay there, i.e. the irregular behaviour is due to  
a chaotic transient rather than a proper chaotic 
attractor.

The survival time on the transient increases dramatically
with increasing $N$ and $\gamma$.
It is hard to decide from numerical results whether the
average survival time $\langle t_s \rangle$ diverges with a 
power law ($\langle t_s \rangle \propto |\gamma -\gamma_d|^{-a}$),
as one usually finds in scenarios where a chaotic transient
becomes a chaotic attractor \cite{Ott:Chaos}, or whether
$\langle t_s \rangle$ increases exponentially with $\gamma$.
In either case, the system shows chaotic behaviour for
sufficiently long times to get stable numerical results -
for example, for $N=20$ and $\gamma=7.0$, the average 
survival time is on the order of $10^6$ steps.
 
If $\gamma$ is larger than some critical value that
depends on $N$, the chaotic transient can eventually
end in a cycle that is related to a 
possible cycle of the discrete CBG. By ``related'' we mean that 
$S^t$ in the CSG is very close to $\pm 1$ and that clipping
the sequence to the nearest value of $\pm1$ would give
the equivalent attractor of the CBG. More different cycles  
become stable with higher $\gamma$; however, the cycle
lengths are usually of order $2N$ -- short cycles are
apparently more likely to become stable than ones whose 
length is of order $2^N$.  

\begin{figure}
\epsfxsize= 0.95\columnwidth
  \epsffile{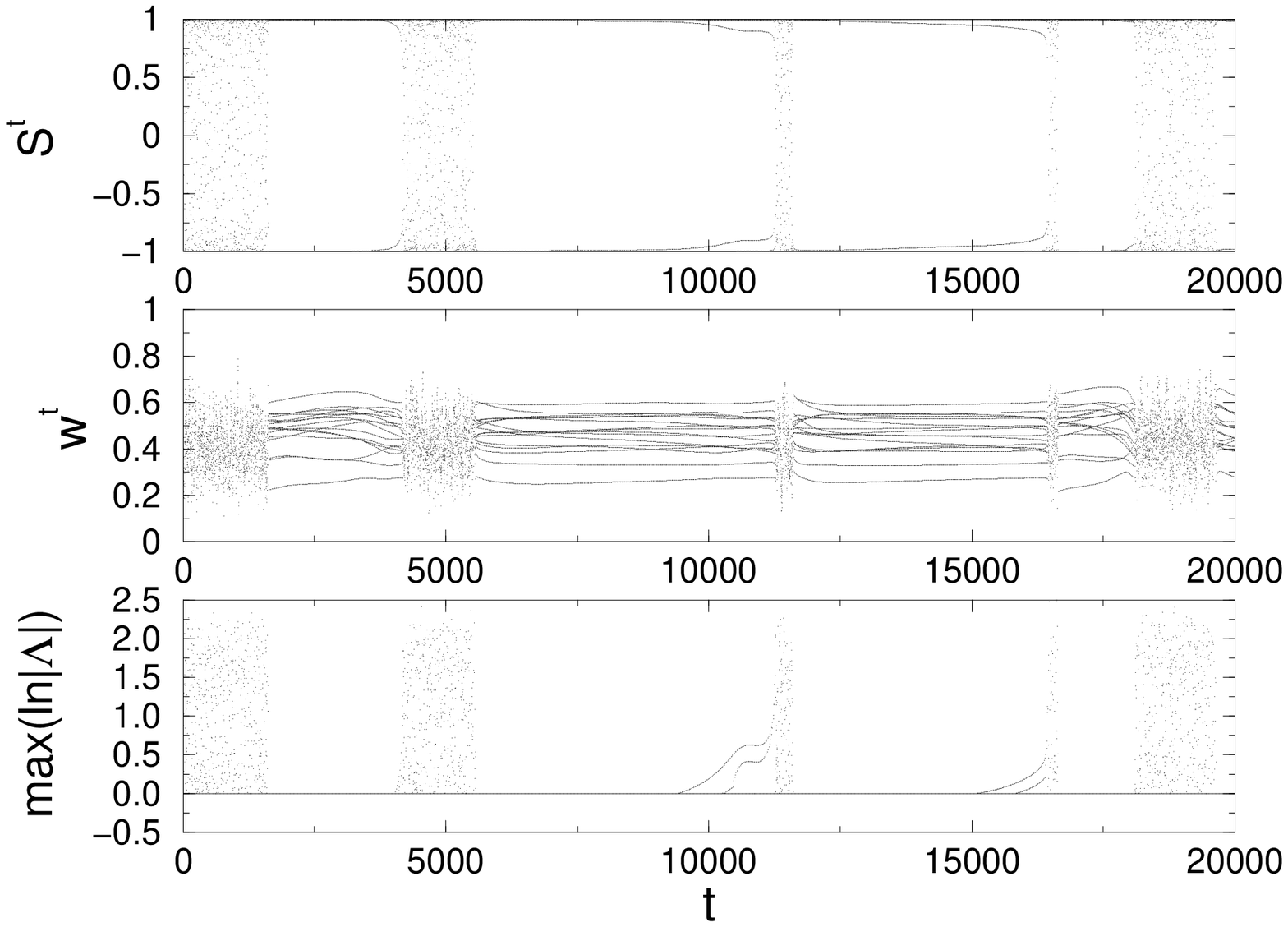}
\caption{Example of intermittent behaviour for $N=6$, $\gamma =8$.
From top to bottom: output $S^t$, norm of weights $w^t$ and 
largest ``one-step Lyapunov exponent'' $\max(\ln(|\Lambda|)$
(see Section \ref{SEC-lyapunov}).}
 \label{CSG-interm}
\end{figure}

At amplifications $\gamma$ slightly below the 
lowest $\gamma$ for which the first cycle becomes stable
for for a given $N$, 
intermittent behaviour is observed: both $S^t$ and
$w^t$ stay near a cycle for an extended number of 
steps (typically several thousand steps for $N=6$) before returning
to chaotic behaviour for a similar time. An example of this 
is given in Fig. \ref{CSG-interm}.

\subsection{Stability and Lyapunov exponents}
\label{SEC-lyapunov}
The term 'chaotic' was used in Sec. \ref{CBG-cyc}
to describe the irregular time series generated
by the CBG. We will now show that the system 
is in fact chaotic in the strict sense.
  
The sensitivity of trajectories of the map (\ref{CSG-upS}),
(\ref{CSG-upw}) to small changes in the initial
conditions  
can be tested by calculating the eigenvalues of the 
Jacobi matrix  
\be
\mathbf{M}^t = \left ( \begin{array}{cc}
    \pdd{x_i^{t^1}}{x_j^{t}} & \pdd{x_i^{t+1}}{w_j^t} \\
    \pdd{w_i^{t+1}}{x_j^{t}} & \pdd{w_i^{t+1}}{w_j^t}
\end{array} \right ).
\ee
This is to be understood as a $2N \times 2N$ matrix with
indices $i$ and $j$ running from 1 to $N$.
The entries of this matrix are of the following form:
\bea
\pdd{x_1^{t+1}}{x_j^t} &=& 
   - \beta w_j^t \frac{2}{\sqrt{\pi}} \exp (-\beta^2 h^2); \nonumber \\
\pdd{x_i^{t+1}}{x_j^t} &=& 
  \delta_{i-1,j}  \mbox{ for } i =2,\ldots,N; \nonumber \\
\pdd{x_1^{t+1}}{w_j^t} &=& 
  - \beta x_j^t \frac{2}{\sqrt{\pi}} \exp(-\beta^2 h^2); \nonumber \\
\pdd{x_i^{t+1}}{w_j^t} &=& 
  0  \mbox{ for } i =2,\ldots,N; \nonumber \\
\pdd{w_i^{j+1}}{x_j^t} &=&
  - \frac{\eta}{N} \erf(\beta h) \delta_{i,j} 
  - \frac{\eta}{N} \beta w_j^t x_i^t \frac{2}{\sqrt{\pi}} 
  \exp (- \beta^2 h^2);  \nonumber \\
\pdd{w_i^{t+1}}{w_j^t} &=& \delta_{i,j} 
  - \frac{\eta}{N} \beta x_j^t x_i^t \exp(-\beta^2 h^2). \label{CSG-jacobi}
\eea
If $|\beta h|$ is large and the transfer function
is saturated, the exponential terms in Eq. (\ref{CSG-jacobi})
are negligible. In that case, the upper left section of $\mathbf{M}$ 
is occupied only on the first lower off-diagonal, the lower right
section is the $N\times N$ unity matrix. Since the upper right 
section is identically 0, the lower left part does not
enter into the calculation of the eigenvalues either.
 
This simplified matrix has $N$ eigenvalues 
$\Lambda =0$ and $N$ eigenvalues 
$\Lambda =1$. The eigenvectors of the latter span the
space of weight vectors, where small changes to $\vw^t$  
are transferred unmodified to $\vw^{t+1}$. The eigenvalues
$\Lambda = 0$ all have the same eigenvector, whose
only nonvanishing component is $x_N$, the component of
the sequence vector that is rotated out at $t+1$.  
This means that the eigenvectors do not span the 
whole space and that thus the eigenvalues are not a
reliable measure of the propagation of a disturbance 
in the system. 

If $|\beta h|$ is small enough for the exponential 
terms to have an appreciable effect, the effect on the
eigenvalues is not easy to calculate. By using values
of $\x$ and $\vw$ taken from a run of the simulation
and numerically calculating the eigenvalues, we find that
typically one of the $\Lambda =0$ eigenvalues is changed drastically
and may have an absolute value $|\Lambda| >1$.
This corresponds to a strong susceptibility of the newly generated
sequence component $S_1$ on small changes in $\vw$ or $\x$.
The other eigenvalues only undergo small corrections,
corresponding to the feedback of the new component
to the weights. 

During the regular phases of intermittent
behaviour, the largest eigenvalues of the one-step
matrix are significantly smaller than during the 
chaotic bursts (see Fig. \ref{CSG-interm}) --
corresponding to sequence values that are close to
$S=\pm 1$, and thus a nearly saturated transfer function. 

\begin{figure}
\epsfxsize= 0.90\columnwidth
  \epsffile{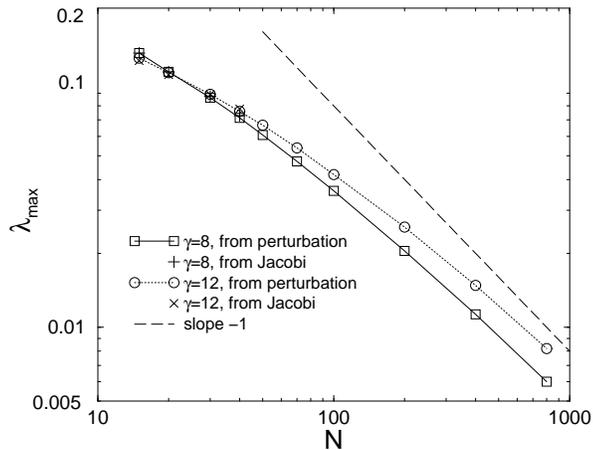}
\caption{Lyapunov exponents measured from the time development 
of perturbations and from iterating the Jacobi matrix 
(Eq. (\ref{CSG-lyapdef})), for
$\gamma = 8$ and $\gamma =12$.}
 \label{CSG-lyapm}
\end{figure}

To find the Lyapunov exponents of the map (see e.g.\cite{Lam:Nonlin}), 
it is necessary to consider the development of a small
perturbation over a long time, i.e. calculate the
eigenvalues $\Lambda^T_i$ of $\prod_{t=1}^{T}\mathbf{M}^t$
(of course, the trajectory is determined using the 
full nonlinear map).
The Lyapunov exponents are then defined as 
\be
\lambda_i = \lim_{T\rightarrow \infty} 
(1/T)\ln |\Lambda^T_i| . \label{CSG-lyapdef}
\ee
The straightforward calculation of the product of
Jacobi matrices brings many numerical problems
which can be eliminated by applying a Gram-Schmidt
orthonormalization procedure to the columns
of the product matrix in regular distances,
as described in \cite{Wolf:Lyapunov}. With this
procedure, it is possible to average over $T>100N$
and get numerically stable results. The largest 
Lyapunov exponent is displayed in Fig. (\ref{CSG-lyapm}).
Typically, there are $N/2$ positive exponents.

The Kaplan-Yorke conjecture \cite{Kaplan:Conjecture} states that
there is a connection between the dimension $D$ of a
attractor of a map and the spectrum of Lyapunov exponents,
which are here assumed to be ordered 
($\lambda_1\geq \lambda_2\geq\ldots\geq \lambda_{2N}$):
 \be 
D_{KY} = k + \sum_{i=1}^{k} \lambda_i / |\lambda_{k+1}|, \label{KY-conj}
\ee 
where $k$ is the value for which $\sum_{i=1}^k \lambda_i > 0$
and  $\sum_{i=1}^{k+1} \lambda_i < 0$. Applying this to 
the spectrum of exponents derived from (\ref{CSG-lyapdef})
gives an average attractor dimension between $1.1 N$ and $1.2N$,
slightly depending on $\gamma$.

An alternative method for measuring the largest 
Lyapunov exponent is to start two trajectories
with infinitesimally different initial conditions,
and propagate both of them using the nonlinear map.   
In regular intervals, measure the distance between
the trajectories, store it, and reset the distance
to the initial value while keeping the direction
of the distance vector. The advantage of this method
is that it requires only $O(N)$ calculations per time
step, rather than $O(N^2)$ like the previous way,
allowing to go to much higher $N$. 

The results for $\lambda_1$ are also 
displayed in Fig. \ref{CSG-lyapm}: the values
gained by the two methods agree well within the 
numerical errors. For large $N$, $\lambda_{max}$
decreases with $1/N$, i.e. perturbations 
grow on the $\alpha$-timescale of online learning.

\section{Summary}
In this paper, we have studied the properties of a 
time sequence generated by a perceptron which 
learns the opposite of its own prediction.
In the case of the simple perceptron, some 
properties are accessible analytically through
the application of online learning techniques
and through the connection between the weights
and the autocorrelation function of the sequence.
The distribution of learning rates among the 
weight components has a decisive influence on 
the statistical properties of the generated
sequence and allows for sequences with 
a wide variety of autocorrelation shapes.

Due to the discrete nature of the sequence and
the learning algorithm, cycles of the system
are inevitable. We find that their typical 
length, as well as that of the transient, grow 
exponentially with the system size $N$.

A histogram of substrings of the generated sequence reveals
that the sequence has significant deviations from 
randomness, although the deviations decrease
with increasing $N$.

Replacing the sign function in the update
rule by a continuous sigmoidal function changes 
many of these results. A vanishing solution
now becomes possible; only for sufficiently
large values of the rescaled amplification $\gamma$
can nontrivial solutions survive. The critical
$\gamma_c$ can be calculated in a mean-field
online learning calculation.

Since both sequence and weights are now 
continuous, cycles vanish for low values
of $\gamma$, and the trajectory is a chaotic sequence.
The largest Lyapunov exponent scales like 
$1/N$ for large $N$; the spectrum of Lyapunov
exponents suggests high-dimensional chaos.

At least some cycles of the CBG reemerge as stable fixed
points of the CGS above a critical $\gamma$ that is
different for each attractor.
Slightly below the smallest critical
$\gamma$ for a given $N$, intermittent behaviour
is observed. 

Compared to the behaviour of sequence-generating
perceptrons with fixed weights, the sequence
generated with changing weights shows more
complex behaviour: longer cycles, more randomness,
chaotic as opposed to quasiperiodic behaviour.
It seems likely that this tendency also holds for 
other algorithms in which the weights keep changing.

\section{Acknowledgements}
R.M., W.K. and I.K. are grateful for financial support by
the German-Israeli Foundation. We thank Stephan Mertens,
Avner Priel, and Andreas Engel 
for discussions, know-how and ideas.

\end{multicols}
\end{document}